# A Class of topological nodal rings and its realization in carbon networks

Yan Gao[1], Yuanping Chen[1]*, Yuee Xie[1], Po-Yao Chang[2], Marvin L. Cohen[3], and Shengbai Zhang[4]

[1]*School of Physics and Optoelectronics, Xiangtan University, Xiangtan, 411105, Hunan, China*

[2]*Center for Materials Theory, Rutgers University, Piscataway, New Jersey, 08854, USA*

[3]*Department of Physics, University of California at Berkeley, and Materials Sciences Division, Lawrence Berkeley National Laboratory, Berkeley, California 94720, USA*

[4]*Department of Physics, Applied Physics, and Astronomy Rensselaer Polytechnic Institute, Troy, New York, 12180, USA*

## Abstract

Topological nodal rings can be classified into three types according to the slopes in their energy dispersion. The first two are made of type-I and II nodal points, respectively, while the third is made of both. In carbon networks, all three types can exist. Under strain, phase transitions from a topological metal to a semiconductor, take place, and at the transition points, these nodal rings shrink into type-I, II, and III semi-Dirac points. These topological features exhibit diverse electron-hole pocket patterns and Landau levels, which give rise to exotic transport properties.

Corresponding author: chenyp@xtu.edu.cn

The study of topological semimetals/metals (TMs) are at the forefront of research in the materials and physical sciences[1-4]. With the discoveries of exotic band crossings near the Fermi level between two[5,6] or multiple bands[7-9], new topological concepts have emerged, for example, from the Dirac point[10,11] or Weyl point[12,13], to a nodal line[14-16], and from type-I to type-II topological points[17-19]. Each new finding expands the class of topological systems often with new physical properties. Therefore, searches for new classes of topological matter have been a recent focus[20,21].

Adjusting band dispersion is an effective way to create new topological systems. The distinction between type-I and type-II Weyl points serves as a good example[22,23]. Both of these result from crossing points between linear bands having qualitatively different slopes. The conduction and valence bands of a type-II point have the same signs, while those of a type-I point have opposite signs[24,25]. Unlike relativistic high-energy physics, the violation of Lorentz invariance for type-II Weyl points reveals the unnecessary restriction of Lorentz symmetry in condensed matter systems[26]. Nevertheless type-I and II Weyl points share the same topological characteristics, and the two can be transformed from one to the other by a Lifshitz transition[27,28], and their Fermi surfaces, which are crucial to magnetic properties and carrier transport, can be drastically different[29,30].

Because of its rich bonding chemistry, a carbon network provides a natural platform in the search for new topological matter[31-33], since carbon has many forms of allotropes ranging from one (1D) to three (3D)dimension[34-40]. Here, the presence of topological properties originates from graphene as a two-dimensional (2D) topological matter[41], with nanoribbons serving as building blocks for the carbon allotropes. The atomic structures of the carbon networks exhibit a wide variety of topological properties ranging from Weyl/Dirac points, nodal loops, to nodal surfaces[42,43]. 3D carbon honeycomb as one kind of carbon networks has been synthesized successfully[38]. In addition, the band dispersion of the networks can be tuned by a geometry change as well as by applying an external strain[44,45].

In this paper, we classify topological nodal rings based on their energy dispersions and demonstrate this classification in carbon networks. We find that nodal rings in such networks belong to three types: type-I is made of two bands of opposite-signed slopes; type-II is made of two bands with slopes of the same-sign; and an anisotropic type-III is made of bands having both the same- and opposite-sign slopes in two different directions, with qualitatively different electron and hole pockets (EHP), as well as Landau levels (LLs). Under strain, these rings shrink into specific types of semi-Dirac points. Tight binding (TB) and $k \cdot p$ models are constructed to unveil the underlying physics of the nodal rings and semi-Dirac points.

Our first-principles calculations were based on the density functional theory within the PBE approximation[46] for the exchange-correlation energy. The core-valence interactions were described by the projector augmented-wave (PAW) potentials[47], as implemented in the VASP code[48]. Plane waves with a kinetic energy cutoff of 600 eV were used as the basis set. The calculations were carried out in periodic super cells[49]. The atomic positions were optimized by using the conjugate gradient method, in which the energy convergence criterion between two consecutive steps was set at $10^{-6}$ eV. The maximum allowed force on the atoms is $10^{-3}$ eV/Å.

Figures 1(a) and 1(b) show the atomic structures of two graphene networks. Both of them are formed by connecting armchair nanoribbons of widths $m$ and $n$ shown in Fig. 1(c), named as AGNW-($m,n$). The variation of $m$ and $n$ produces a series of carbon networks, and Figs. 1(a) and 1(b) correspond to AGNW-(3,2) and (1,2), respectively. At the connecting lines, the (gray) atoms are $sp^3$ hybridized, while the atoms in the nanoribbons of widths $m$ and $n$, labeled as blue $C_1$ and pink $C_2$ respectively, are $sp^2$ hybridization. Figure 1(d) shows the conventional and primitive cells of AGNW-(1,2), where the armchair chains are along $z$ axis.

The space groups of AGNW-(3,2) and (1,2) are both IMMA, where there is a mirror $M_z$ normal to $z$ axis. The bond lengths are in the range of 1.35 ~ 1.57 Å. Most of them are between those of diamond (1.54 Å) and graphite (1.42 Å). The porous structures here lead to small densities even lower than that of graphite. The calculated cohesive energies $E_{coh}$ indicate that AGNW-(3,2) (-7.67 eV/C) and AGNW-(1,2) (-7.47 eV/C) are metastable allotropes, as they are only 0.1 ~ 0.3 eV/C smaller than that of diamond. To establish the stability, we calculate phonon spectra for these cases, and we don't find any soft modes over the entire Brillouin Zone (BZ). [Details can be seen in Table S1 and Fig. S1 in Supplementary Information (SI)].

It is known that electronic properties of armchair nanoribbons depend on their widths. The band gaps of ribbons with width 3p+2 (p is an integer) are much smaller than others[50], and the bands around the Fermi level are characterized by the $p_z$ orbital. Therefore, the electronic states around the Fermi level in an AGNW are dominated by the nanoribbons of width 3p+2. Both AGNW-(3,2) and (1,2) include nanoribbons of width n = 2. To explore their electronic properties, we only need consider one orbital of the atoms in these nanoribbons. In the primitive cells of AGNW-(3,2) and (1,2), these are the four pink atoms in the nanoribbons of n = 2. Therefore, a 4×4 TB model can be constructed to describe the networks:

$$H = \sum_{<i,j>} \sum_{\mu} t_{ij} e^{-i\boldsymbol{k} \cdot \boldsymbol{d}_{ij}^{\mu}}, \tag{1}$$

where $i, j \in \{1,2,3,4\}$ label the four pink $C_2$ atoms, $d_{ij}^{\mu}$ is the displacement vector directed from atoms j to i, $t_{ij}$ is the hopping energy between atoms i and j, and μ runs over all lattice sites under translation. For $t_{ij}$, we consider five parameters: the intra-interactions $t_1$ and $t_2$ of nanoribbons, and inter-interactions $t_3$, $t_4$ and $t_5$ [see Figs. 1(c-d)]. We note that, in the structures AGNW-(m,2), $t_3$, $t_4$ and $t_5$ vary with m, while all parameters can be tuned by external strains.

Our calculations indicate that the variations of $t_3$, $t_4$ and $t_5$ have drastic effects on the band dispersions. According to the dispersions, the band spectra can be classified into types: type I, II and III [see Figs. S2 in SI]. When $t_2$ is further changed, each type of band spectrum experiences a phase transition from having a topological nodal ring to gap state, and the critical phase is a semi-Dirac point. Figures 2(a-c) show schematically the three types of semi-Dirac points, where the contacting bands are all linear along $k_z$ while they are all quadratic along $k_x$ and $k_y$. The difference of the three types is the slope signs of the quadratic bands. The type-I case in Fig. 2(a) is a standard semi-Dirac point[51,52], where the conduction and valence bands have opposite signs. While type-II and III in Figs. 2(b-c) are two new types of semi-Dirac points. In type-II, the quadratic valence bands have the same slope signs with the conduction bands. While type-III semi-Dirac point exhibits the most remarkable anisotropy: two linear bands along $k_z$, two quadratic bands with same and opposite slope signs along $k_x$ and $k_y$, respectively.

When $t_2$ is increased, the valance and conduction bands cross along the $k_x$ and $k_y$ directions, and then the semi-Dirac points evolve into three types of topological nodal rings [see Figs. S2(g-i) in SI]. The 3D band spectra of the three types of nodal rings are shown schematically in Figs. 2(d-f), respectively. Both type-I and type-II rings can be viewed as crossing lines of two paraboloids. The difference is the former has the opposite opening while the latter has the same. Therefore, a type-I ring is made of type-I Dirac points[53-55], while a type-II ring is made of type-II Dirac points. The type-III ring in Fig. 2(f) is crossed by a paraboloid and a saddle surface, where the Dirac points are type-I along $k_y$ while are type-II along $k_x$. In other words, each point is anisotropic.

Interestingly, the three types of topological nodal rings, as well as the phase transitions mentioned above, can be realized in AGNW-(3,2) and (1,2). Figure 3(a) shows the band structure of AGNW-(3,2). One can find that, along $\Gamma$-X, $\Gamma$-S and $\Gamma$-Y, there are type-I Dirac points around the Fermi level. More Dirac points on the $k_z = 0$ plane link to a type-I ring [green circle in the inset of Fig. 3(a)]. When a 2.3% strain is applied on axis $z$, the band structure is shown in Fig. 3(b), where the conduction and valance bands only contact at $\Gamma$. Here the contact point is a type-I semi-Dirac point. When the strain is larger than 2.3%, a band gap appears. This means that the strain induces a type-I phase transition in AGNW-(3,2).

In Fig. 3(c), the band structure of AGNW-(1,2) is shown. The carbon network is a metal. At $E$=-0.38 eV, one can note that the two bands only contact at $\Gamma$. The contact point is a type-III semi-Dirac point, exhibiting drastic anisotropic properties. The two contacting bands are linear along $k_z$ ($\Gamma$-Z); are quadratic along $k_x$ ($\Gamma$-X) and $k_y$ ($\Gamma$-Y), and their slopes have the same and opposite signs, respectively. The 3D band structure in the inset of Fig. 3(c) clearly shows the difference of dispersions along $k_x$ and $k_y$. When a 2.2% strain is applied on axis $z$, the band structure is shown in Fig. 3(d). There is a type-II Dirac point along $k_x$, while there are type-I Dirac points along $k_y$. These Dirac points make up a type-III ring on the $k_z = 0$ plane, and this is shown in the inset of Fig. 3(d). When a strain -5.0% is applied on axis $z$, the carbon electronic structure changes to a semiconductor. Therefore, a strain along $k_z$ induces a type-III phase transition. In addition, a type-II ring can also be realized in AGNW-(1,2) by applying a strain along axis $x$ [see Fig. S3 in SI].

The band structures in Fig. 3 are projected onto atomic orbitals. One finds that all the bands around the Fermi level originate from the $p_y$ orbitals of $C_2$ atoms. Therefore, it is reasonable to use Eq. (1) to describe AGNW-(3,2) and (1,2). The charge density of state around the Fermi level also illustrates that their electronic properties are dominated by the $p_y$ orbital of $C_2$ atoms [see the insets of Fig. 3(b)].

A $k \cdot p$ model is useful to further explore the physics of the topological nodal rings. Constrained by the IMMA symmetry group and the time reversal symmetry for a spinless system (spin-orbit coupling is negligible for carbon), one obtains the $k \cdot p$ model around $\Gamma$ up to k-quadratic order described by

$$H(\boldsymbol{k}) = \begin{bmatrix} A_1 k_x^2 + B_1 k_y^2 & iCk_z \\ -iCk_z & \Delta + A_2 k_x^2 + B_2 k_y^2 \end{bmatrix},\qquad(2)$$

where $\Delta$ is the gap at $\Gamma$, and the coefficients $A_1$, $B_1$, $A_2$, $B_2$ and $C$ can be determined by fitting to the DFT or TB results. $\Delta = 0$ and $> 0$ present the cases of semi-Dirac points and nodal rings, respectively. The signs of the parameters $A_1$, $B_1$, $A_2$, $B_2$ determine the dispersions of conduction and valance bands along $k_x$ and $k_y$. By changing the signs, in Table1, we find that there are two other sub-types of type-II and one sub-type of type-III, whose 3D band spectra are shown in Fig. S6 in SI. By changing the sign of one parameter, type-III rings turn into type-I or II rings. Therefore, type-III rings can be considered as intermediate phases.

The three types of nodal rings share the same topological characteristics. All of them are protected by the mirror symmetry $M_z$ normal to $z$ axis. The protection can also be inferred from the 1D winding number of $\mathcal{H}(\boldsymbol{k})$ along a close path $\mathcal{L}$ encircling the rings: $N_{\mathcal{L}} = \frac{1}{2\pi i} \oint_{\mathcal{L}} dl \cdot \mathrm{Tr}[Q^{-1}(\boldsymbol{k})\nabla_l Q(\boldsymbol{k})] = \pm 1$ [56]. Therefore, these nodal rings have similar topological surface states.

It is known that there are EHPs in the nodal ring semimetals, because most nodal points locate around rather than on the Fermi level[14,53]. Figures 4(a-c) compare EHPs of type-I, II and III rings, in the case of the Fermi level crosses the nodal ring (the yellow circles are crossing points). The EHPs of a type-I ring are sandwiched between two equi-energy ellipses, and projection of the ring locates between the equi-energy curves. The EHPs of a type-II ring are complementary with those of type-I, which distribute in and out of the two equi-energy ellipses. If the two ellipses in Figs. 4(a) and 4(b) become circles, the type-I EHPs disappear completely while type-II EHPs occupy the whole space. The type-III ring exhibits drastically anisotropic EHPs: the electron pockets distribute in parts of ellipse along $k_y$, while the hole pockets are out the ellipse along $k_x$. The type-I and II EHPs are two extremes: the area of the former is small while the latter is very large. Relatively, type-III EHPs are more changeable since the EHPs can be tuned by the Fermi level. Moreover, they never disappear because type-III ring cannot locate on an equi-energy plane.

The diverse Fermi surface and EHPs patterns imply that the three types of nodal rings exhibit a variety of physical properties. First of all, their transport properties are very different. For example, the transport of a type-I ring is dominated by electrons and holes along directions *y* and *x*, respectively, while the transport properties of type-II and III rings are attributed to a mixture of electrons and holes in both directions. However, the transport properties of a type-II ring is isotropic, while for the type-III case, it is anisotropic and the degree of anisotropy can be tuned by the Fermi level. This difference could provide an effective signature for determining the types of nodal rings experimentally.

Another substantial signature of the three types of nodal rings are LL spectra, as shown in Figs. 4(d-f), which is based on the simplified k · p models in Table I. For the type-I ring, the LLs are

$$E_{\pm}^n = \frac{1}{2}\Delta \pm \frac{1}{2}\sqrt{\Delta^2 + 4[k_z^2 + E_n(E_n - \Delta)]}, \qquad (3)$$

where $E_n = eB(2n + 1)$. At $k_z = 0$, $E_+^n = \Delta - E_n$ and $E_-^n = E_n$. The energies have a uniform distribution with an interval of $2eB$ (see Fig. 4(d)). For the type-II ring, the LLs are

$$E_{\pm}^n = \frac{1}{2}[E_n(1 + \gamma) + \Delta] \pm \frac{1}{2}\sqrt{[E_n(1 + \gamma) + \Delta]^2 + 4[k_z^2 - E_n(\gamma E_n + \Delta)]}. \qquad (4)$$

At $k_z = 0$, $E_+^n = \Delta + \gamma E_n$ and $E_-^n = E_n$. Here the energy levels of $E_+^n$ shrink because the interval is $2\gamma eB$ ($0 < \gamma < 1$). They collapsed to a degenerated value $\Delta$ at the transition point of type-I and type-II ($\gamma = 0$). This means that the phase transition point is also the critical point of variation for the LLs. Figure 4(e) presents the LLs of an ordinary type-II ring ($\gamma = 0.5$). For the type-III ring, a second ordered perturbation theory can be used to calculate the LLs. At $k_z = 0$, $E_+^n = \Delta + \frac{1}{8}E_n[(\delta + 3)^2 - 12]$ and $E_-^n = E_n$. In this case, $E_+^n$ also shrinks because of $0 < \delta < 1$. When $\delta \approx 0.46$, $E_+^n$ collapses completely to the energy $\Delta$. Figure 4(f) presents the LLs spectrum of a type-III nodal ring with $\delta = 0.5$. By comparing Figs. 4(d-f), one can find that type-III ring has the most remarkable collapse effect. Therefore, the materials with type-III ring could be named as degenerate Landau metals.

Beside the electronic and magnetic responses, the three types of nodal rings may reveal various optical properties. For example, the EHPs of type-I and II rings result in strong and weak optical absorptions[57], respectively; while the type-III ring not only can exhibit intermediate variations between them, but also possesses unique optical, optoelectronic and magneto-optical phenomena because of its anisotropic EHPs. In addition, some other physical phenomena[58-60], such as superconducting, CDW and nesting instability may be possible.

In summary, three classes of topological nodal rings are identified and they are predicted to exist in carbon networks. Type-I rings, made of type-I nodal points, have been extensively discussed in the literature. However, little is known about type-II and III rings. Among the three types, type-III is the one that may deserve the most attention because of its maximum anisotropy. Diverse EHPs of the nodal rings exhibit rich and often completely-different patterns, and this provides a great platform to uncover the fundamental electronic, optical and magnetic properties, such as, anisotropy of electron/hole transports, collapse of Landau-level, and tunable optical absorption. Under strain, nodal rings shrink, eventually into type-I, II, and III semi-Dirac points, respectively. Topological protection disappears at the transition point so a band gap opens. Considering that the type-II and III phases can be produced by strain in the same AGNW, all the transitions of physical properties between the two phases can be tuned by external forces. As the AGNWs include many carbon structures, we expect that other unexplored topological properties may be found in this class of carbon networks which should provide new knowledge for three dimensional and lower dimensional carbon systems.

We thank the discussions with David Vanderbilt. YPC and YEX were supported by the National Natural Science Foundation of China (Nos.51376005 and 11474243). PYC was supported by the Rutgers Center for Materials Theory. MLC was supported by National Science Foundation Grant No. DMR-1508412 (theoretical analysis of computer data) and by the Office of Basic Energy Sciences, Materials Sciences and Engineering Division, U.S. Department of Energy under contract No. DE-AC02-05CH11231, within the SP2 program (electronic structure calculation). SBZ was supported by the US Department of Energy (DOE) under Grant No. DESC0002623.

# References

[1]     A. A. Burkov, Nat Mater **15**, 1145 (2016).

[2]     Z. Liu *et al.*, Science **343**, 864 (2014).

[3]     B. Q. Lv *et al.*, Nat. Phys **11**, 724 (2015).

[4]     T. Bzdusek, Q. Wu, A. Ruegg, M. Sigrist, and A. A. Soluyanov, Nature **538**, 75 (2016).

[5]     A. H. Castro Neto, N. M. R. Peres, K. S. Novoselov, and A. K. Geim, Rev. Mod. Phys. **81**, 109 (2009).

[6]     S. M. Young and C. L. Kane, Phys. Rev. Lett. **115**, 126803 (2015).

[7]     Z. Zhu, G. W. Winkler, Q. Wu, J. Li, and A. A. Soluyanov, Phys. Rev. X **6**, 031003 (2016).

[8]     B. J. Wieder, Y. Kim, A. M. Rappe, and C. L. Kane, Phys. Rev. Lett. **116**, 186402 (2016).

[9]     B. Bradlyn, J. Cano, Z. Wang, M. G. Vergniory, C. Felser, R. J. Cava, and B. A. Bernevig, Science **353**, aaf5037 (2016).

[10]     J. A. Steinberg, S. M. Young, S. Zaheer, C. L. Kane, E. J. Mele, and A. M. Rappe, Phys. Rev. Lett. **112**, 036403 (2014).

[11]     H. Li, H. He, H. Z. Lu, H. Zhang, H. Liu, R. Ma, Z. Fan, S. Q. Shen, and J. Wang, Nat Commun **7**, 10301 (2016).

[12]     L. Lu, Z. Wang, D. Ye, L. Ran, L. Fu, J. D. Joannopoulos, and M. Soljačić, Science **349**, 622 (2015).

[13]     B. Lv *et al.*, Phys. Rev. X **5**, 031013 (2015).

[14]     R. Li, H. Ma, X. Cheng, S. Wang, D. Li, Z. Zhang, Y. Li, and X. Q. Chen, Phys. Rev. Lett. **117**, 096401 (2016).

[15]     Y. Kim, B. J. Wieder, C. L. Kane, and A. M. Rappe, Phys. Rev. Lett. **115**, 036806 (2015).

[16]     A. Burkov, M. Hook, and L. Balents, Phys. Rev. B **84**, 235126 (2011).

[17]     S.-Y. Xu *et al.*, Nat. Phys **11**, 748 (2015).

[18]     M. Udagawa and E. J. Bergholtz, Phys. Rev. Lett. **117**, 086401 (2016).

[19]     Z. Wang, D. Gresch, A. A. Soluyanov, W. Xie, S. Kushwaha, X. Dai, M. Troyer, R. J. Cava, and B. A. Bernevig, Phys.Rev.lett **117**, 056805 (2016).

[20]     J. Jiang *et al.*, Nat. Commun **8**, 13973 (2017).

[21]     G. Autes, D. Gresch, M. Troyer, A. A. Soluyanov, and O. V. Yazyev, Phys. Rev. Lett. **117**, 066402 (2016).

[22]     S.-Y. Xu *et al.*, Science **349**, 613 (2015).

[23]     Y. Wang *et al.*, Nat. Commun **7**, 13142 (2016).

[24]     G. Chang *et al.*, Sci. Adv **2**, e1600295 (2016).

[25]    K. Koepernik, D. Kasinathan, D. Efremov, S. Khim, S. Borisenko, B. Büchner, and J. van den Brink, Phys. Rev. B **93**, 201101 (2016).

[26]    S.-Y. Xu *et al.*, Sci. Adv **3**, e1603266 (2017).

[27]    G. E. Volovik, Low. Temp. Phys **43**, 47 (2017).

[28]    A. Cortijo, D. Kharzeev, K. Landsteiner, and M. A. H. Vozmediano, Phys. Rev. B **94**, 241405 (2016).

[29]    Z.-M. Yu, Y. Yao, and S. A. Yang, Phys. Rev. Lett. **117**, 077202 (2016).

[30]    A. Tamai *et al.*, Phys. Rev. X **6**, 031021 (2016).

[31]    Y. Lin, Z. Zhao, T. A. Strobel, and R. E. Cohen, Phys. Rev. B **94**, 245422 (2016).

[32]    H. Weng, Y. Liang, Q. Xu, R. Yu, Z. Fang, X. Dai, and Y. Kawazoe, Phys. Rev. B **92**, 045148 (2015).

[33]    Y. Cheng, X. Feng, X. Cao, B. Wen, Q. Wang, Y. Kawazoe, and P. Jena, Small **13**, 1602894 (2017).

[34]    M. Ouyang, J.-L. Huang, C. L. Cheung, and C. M. Lieber, Science **292**, 702 (2001).

[35]    K. S. Novoselov, A. K. Geim, S. V. Morozov, D. Jiang, Y. Zhang, S. V. Dubonos, I. V. Grigorieva, and A. A. Firsov, Science **306**, 666 (2004).

[36]    C. Zhong, Y. Xie, Y. Chen, and S. Zhang, Carbon **99**, 65 (2016).

[37]    J. T. Wang, H. Weng, S. Nie, Z. Fang, Y. Kawazoe, and C. Chen, Phys. Rev. Lett. **116**, 195501 (2016).

[38]    N. V. Krainyukova and E. N. Zubarev, Phys. Rev. Lett. **116**, 055501 (2016).

[39]    Y. Gao, Y. Chen, C. Zhong, Z. Zhang, Y. Xie, and S. Zhang, Nanoscale **8**, 12863 (2016).

[40]    Y. Chen, Y. Y. Sun, H. Wang, D. West, Y. Xie, J. Zhong, V. Meunier, M. L. Cohen, and S. B. Zhang, Phys. Rev. Lett. **113**, 085501 (2014).

[41]    K. S. Novoselov, A. K. Geim, S. V. Morozov, D. Jiang, M. I. Katsnelson, I. V. Grigorieva, S. V. Dubonos, and A. A. Firsov, Nature **438**, 197 (2005).

[42]    Y. Chen, Y. Xie, S. A. Yang, H. Pan, F. Zhang, M. L. Cohen, and S. Zhang, Nano. Lett **15**, 6974 (2015).

[43]    C. Zhong, Y. Chen, Y. Xie, S. A. Yang, M. L. Cohen, and S. B. Zhang, Nanoscale **8**, 7232 (2016).

[44]    R. Fei, V. Tran, and L. Yang, Phys. Rev. B **91**, 195319 (2015).

[45]    C. Zhong, Y. Chen, Z. M. Yu, Y. Xie, H. Wang, S. A. Yang, and S. Zhang, Nat. Commun **8**, 15641 (2017).

[46]    J. P. Perdew, K. Burke, and M. Ernzerhof, Phys. Rev. Lett. **77**, 3865 (1996).

[47]    G. Kresse and D. Joubert, Phys. Rev. B **59**, 1758 (1999).

[48]    G. Kresse and J. Furthmüller, Comput. Mater. Sci **6**, 15 (1996).

[49]    M. L. Cohen, M. Schlüter, J. R. Chelikowsky, and S. G. Louie, Phys. Rev. B **12**, 5575 (1975).

[50]    V. Meunier, A. G. Souza Filho, E. B. Barros, and M. S. Dresselhaus, Rev. Mod. Phys. **88**, 025005 (2016).

[51]    V. Pardo and W. E. Pickett, Phys. Rev. Lett. **102**, 166803 (2009).

[52]    S. Banerjee, R. Singh, V. Pardo, and W. Pickett, Phys. Rev. Lett. **103**, 016402 (2009).

[53]    R. Yu, H. Weng, Z. Fang, X. Dai, and X. Hu, Phys. Rev. Lett. **115**, 036807 (2015).

[54]    L.-K. Lim and R. Moessner, Phys. Rev. Lett. **118**, 016401 (2017).

[55]    G. Bian *et al.*, Phys. Rev. B **93**, 121113 (2016).

[56]    G. Bian *et al.*, Nat. Commun **7**, 10556 (2016).

[57]    S. Li, Z.-M. Yu, Y. Liu, S. Guan, S.-S. Wang, X. Zhang, Y. Yao, and S. A. Yang, arXiv preprint arXiv:1705.02076  (2017).

[58]    D. LeBoeuf *et al.*, Nature **450**, 533 (2007).

[59]    T. Kidd, T. Miller, M. Chou, and T.-C. Chiang, Phys. Rev. Lett. **88**, 226402 (2002).

[60]    I. Mazin, D. J. Singh, M. Johannes, and M.-H. Du, Phys. Rev. Lett. **101**, 057003 (2008).

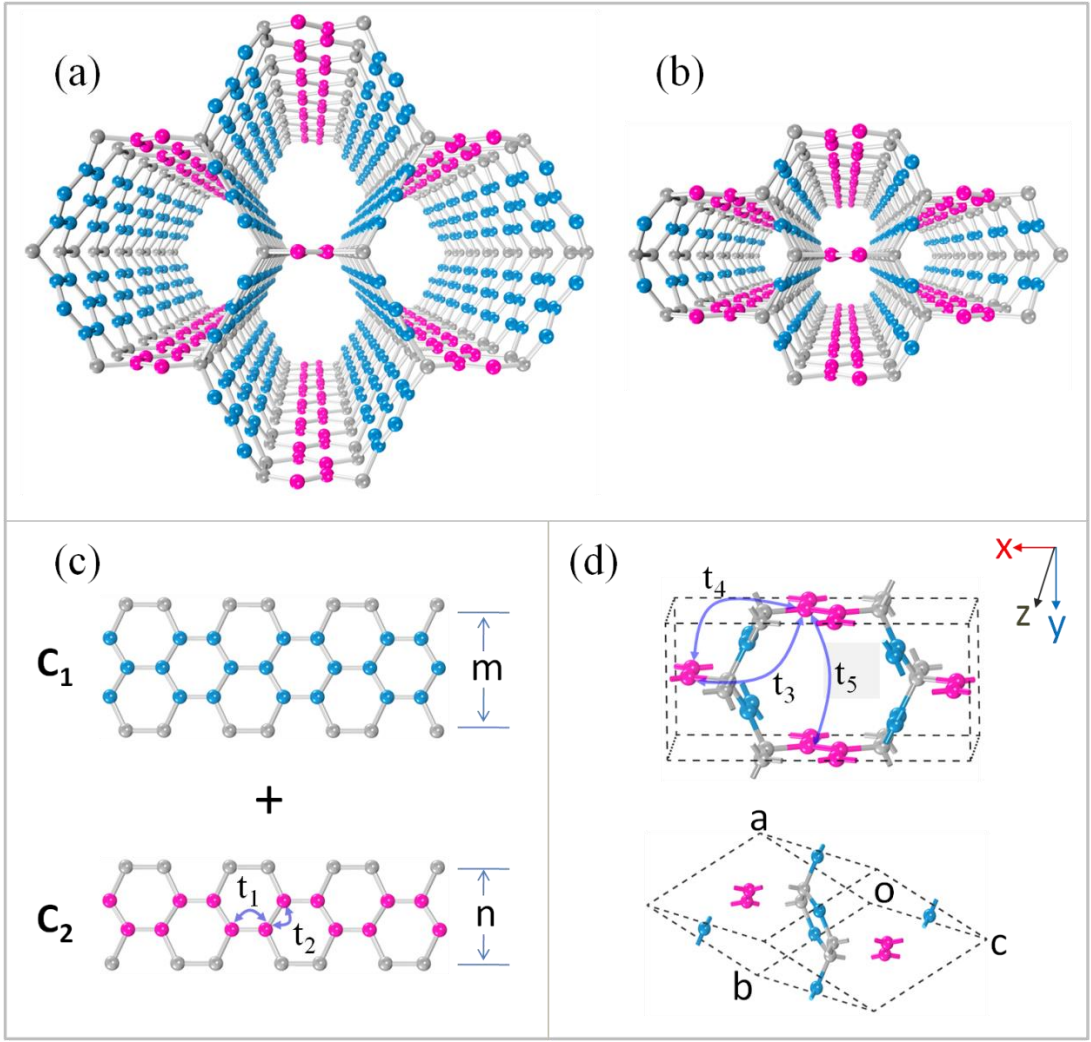

Figure 1. Atomic structures of (a) AGNW-(3,2) and (b) AGNW-(1,2). The two structures belong to the series of AGNW-(m,n), formed by connecting two kinds of armchair nanoribbons in (c) with widths m and n, respectively. The blue (C1) and pink (C2) atoms are $sp^2$ atoms while the gray atoms are $sp^3$ atoms linking the $sp^2$ atoms. (d) Conventional (top) and primitive (bottom) cells of AGNW-(1,2). $t_1$, $t_2$ in (c) are hopping energies in the nanoribbons of C2 atoms, while the $t_3$, $t_4$ and $t_5$ in (d) are hopping energies between the nanoribbons.

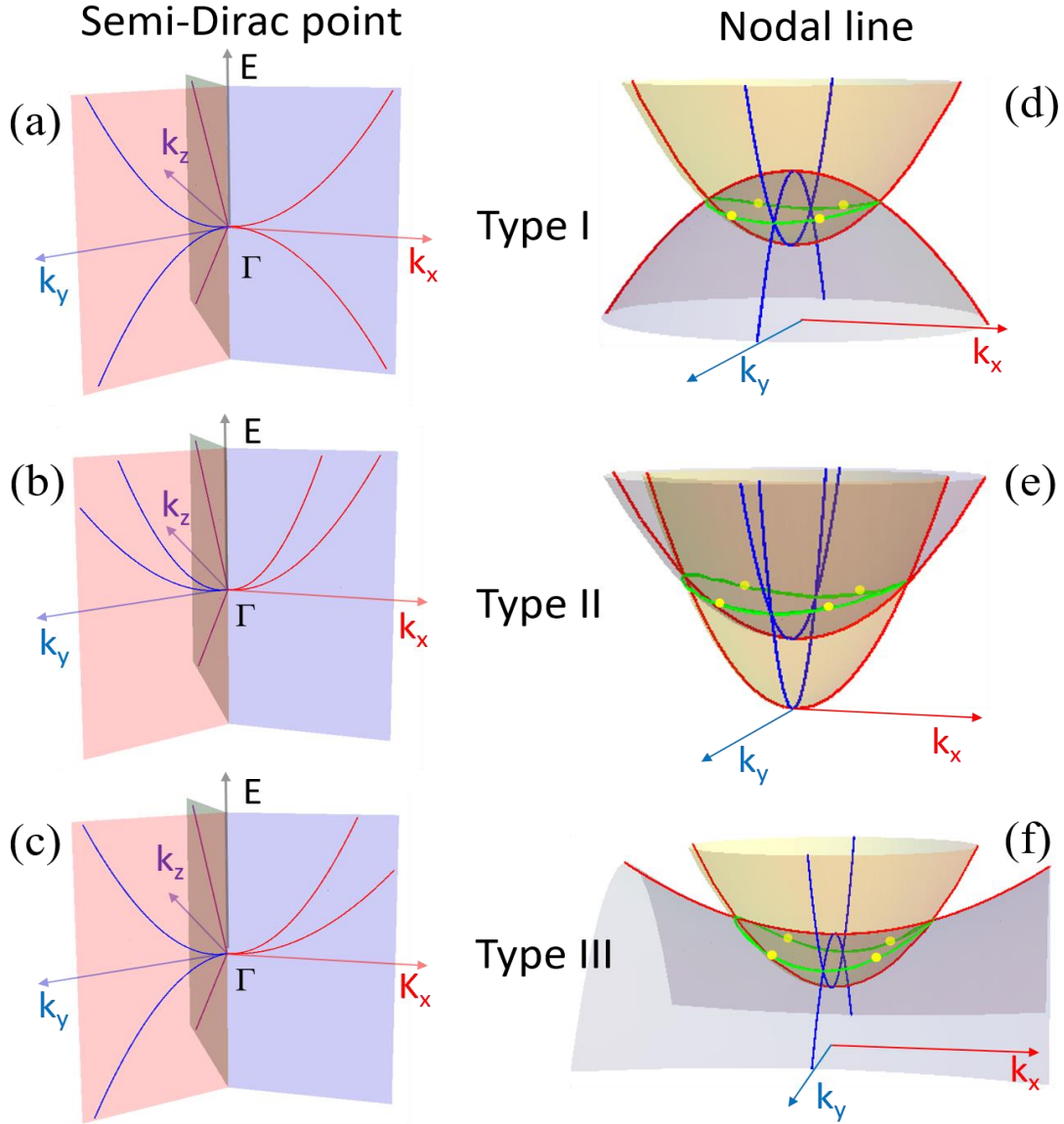

Figure 2. Three types of semi-Dirac points and topological nodal rings. (a) Type-I, (b) type-II and (c) type-III semi-Dirac points. The red, blue and olive lines represent energy bands along kx, ky and kz, respectively. The 3D band spectra of (d) Type-I, (e) type-II and (f) type-III topological nodal rings. The green circles represent nodal rings. The four yellow points represent crossing points of the nodal rings and the Fermi level (or an equi-energy surface).

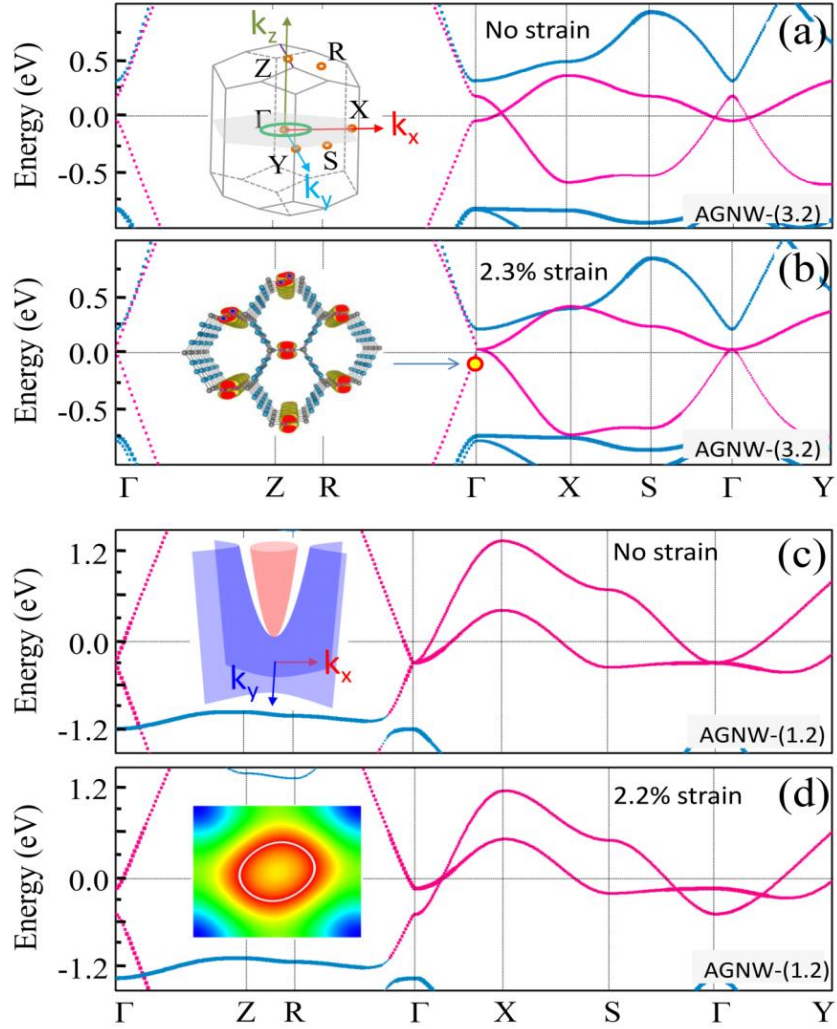

Figure 3. Band structures of AGNW-(3,2) with (a) no strain and (b) a 2.3% strain along axis z. The insets in (a): the first BZ, where the green circle represents location of the nodal ring; and in (b): a charge density of state around the semi-Dirac point. Band structures of AGNW-(1,2) with (a) no strain and (b) a 2.2% strain along axis z. The insets in (c): 3D band spectrum of the semi-Dirac point; and in (d): contour plot of type-III nodal ring on $k_z = 0$ plane. All the bands are projected to atomic orbitals. The pink and blue color bands correspond to $p_y$ orbital of C2 atoms and $p_{x/y}$ orbital of C1 atoms, respectively.

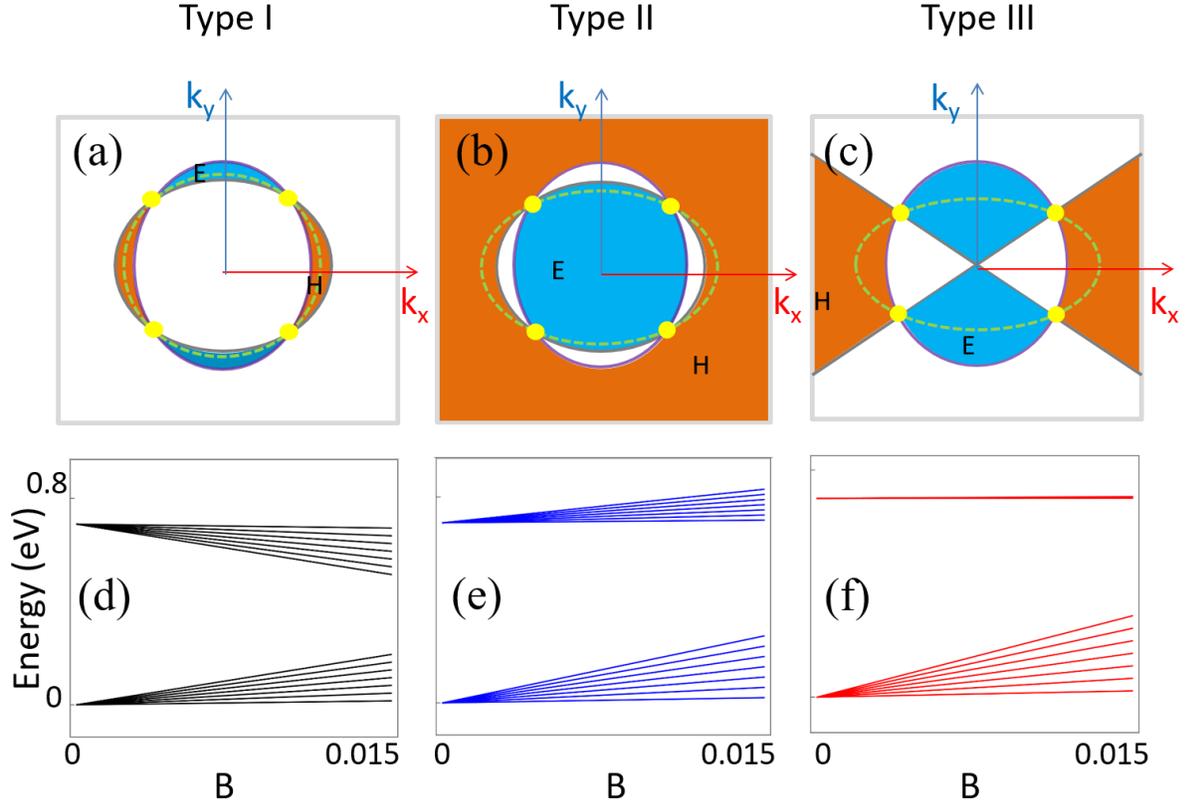

Figure 4. EHPs around (a) type-I, (b) type-II and (c) type-III nodal rings. The light blue regions label electron pockets (E), while the orange regions label hole pockets (H). The four yellow points are nodal points on the nodal rings, which correspond to the four yellow points in Figs. 2(d-f). The green dotted circles are projections of the nodal lines in Figs. 2(d-f). LLs as a function of magnetic field B (in unit of T) ($k_z = 0$, $\Delta = 0.7$ eV) for (d) type-I (e) type-II ($\gamma$=0.5) and (f) type-III ($\delta$=0.5) nodal rings.

Table 1. The signs of the parameters in Eq. (2) for the three types of topological nodal rings, where type-II and type-III have 3 and 2 sub-types, respectively. The last column gives the values of the parameters for simplified models of the three types of rings.

| Type | Sub type | $A_1$ | $B_1$ | $A_2$ | $B_2$ | Simplified model (C = 1) |
|------|----------|-------|-------|-------|-------|--------------------------|
| I | 1 | + | + | - | - | $A_1 = B_1 = -A_2 =- B_2 = 1$ |
| II | 1 | + | + | + | + | $A_1 = B_1 = 1$, $A_2 = B_2 = \gamma$ $(0 < \gamma < 1)$ |
| | 2 | - | - | - | - | |
| | 3 | + | - | + | - | |
| III | 1 | + | + | + | - | $A_1 = B_1 = -B_2 = 1$, $A_2 = \delta$ $(0 < \delta < 1)$ |
| | 2 | + | - | - | - | |